\documentclass[a4paper,11pt]{article}
\pdfoutput=1 

\usepackage{jcappub} 

\usepackage[utf8]{inputenc}
\usepackage[T1]{fontenc} 

\usepackage{graphicx}	
\usepackage{amsmath}	
\usepackage{url}

\usepackage{newtxtext,newtxmath}

\usepackage{subcaption}
\usepackage{adjustbox}
\DeclareUnicodeCharacter{2212}{-}
\usepackage{gensymb}
\usepackage{tabularx}

\title{\boldmath Cherenkov Telescope Array Observatory sensitivity to heavy Galactic Cosmic Rays and the shape of particle spectrum}

\author[a,1]{C. Dubos,\note{Corresponding author.}}
\author[a,b]{P. Sharma,}
\author[a,2]{S. Patel\note{Also at Helmholtz-Zentrum Berlin für materialien und energie.}}
\author[a]{and T. Suomij\"{a}rvi}

\affiliation[a]{Universit\'{e} Paris-Saclay, CNRS/IN2P3, IJCLab,\\91405 Orsay, France}
\affiliation[b]{Universit\'{e} de Strasbourg, CNRS, Observatoire astronomique de Strasbourg, UMR 7550,\\F-67000 Strasbourg, France}

\emailAdd{coline.dubos@ijclab.in2p3.fr}
\emailAdd{pooja.sharma@unistra.fr}
\emailAdd{tiina.suomijarvi@ijclab.in2p3.fr}

\abstract{The origin of Galactic Cosmic Rays (GCRs) and the potential role of Supernova Remnants (SNRs) as cosmic-ray (CR) accelerators remain subjects of ongoing debate. To shed more light on this topic, we have studied the spectral \textcolor{red}{shapes} of two SNRs, RX J1713.7-3946 and HAWC J2227+610, performing simulations for the Cherenkov Telescope Array Observatory (CTAO). 
The previous multi-wavelength (MWL) analysis on these two sources showed an important hadronic contribution at high energies. The interaction of the GCRs accelerated by the SNRs with the medium around the accelerator leads to a process of pion decay (PD) that produces gamma-rays ($\gamma$-rays). These emissions, detectable by CTAO, offer an indirect means of pinpointing the CR source. \\
Two scenarios have been considered: the contribution of heavy CRs and different cut-off sharpnesses ($\beta$) of the particle spectra. 
The simulations were performed by using different CR composition distributions (protons, CNO, Fe) and different sharpness values ranging from $\beta$=0.5 to $\beta$=1.5.\\ The results show that, in the cases studied here, CTAO will increase the sensitivity to the spectral shape of $\gamma$-rays. This allows us to distinguish protons from heavy CRs and obtain information on $\beta$ values and therefore on different acceleration scenarios.}


\begin{document}
\label{firstpage}
\maketitle
\flushbottom

\section{Introduction}
\label{intro}

For over a century, direct and indirect observations have provided detailed information on the energy spectrum of CRs. This spectrum spans an impressive range of 12 orders of magnitude in energy and 32 in flux. 
A pronounced softening of the CR spectrum is observed at around $3\times10^{15}$\,eV. This so-called knee is attributed to light CR primaries (such as protons) \cite{ANTONI20051}. A less pronounced second knee attributed to heavy primaries (such as Fe), is observed at around $9\times10^{16}$\,eV \cite{HORANDEL2004241, hillas2006cosmic} and a mixed composition is observed between these two energies.\\ 



The CRs below $10^{18}$\,eV are believed to be accelerated by Galactic sources. Above this energy, the sources are shown to be extragalactic \cite{2017}. The identification of Galactic sources that can accelerate CRs up to the knee energies, also known as PeVatrons, remains an active field of research.
Supernova (SN) explosions are among the most violent high-energy phenomena in our Galaxy and their remnants (SNRs) emerge as one of these potential Galactic sources with an energy budget sufficient to accelerate CRs up to the PeV energy \cite{Cristofari_2021} (and references therein). Approximately 10-20 per cent 
of the kinetic energy from a SN explosion is converted into accelerated particles through diffusive shock acceleration (DSA) \cite{1983RPPh...46..973D, Damiano_2011}. If one supposes three SN explosions per century in the Milky Way, the energy budget would be sufficient to reproduce the CR spectrum.\\

After being accelerated by SNR shock waves, CRs \textcolor{red}{could} escape from the remnant and interact with the surrounding medium. This can lead to emission of high-energy $\gamma$-rays through the inelastic production of neutral pions $\pi^0$ in proton-proton collisions: 
\begin{equation}
\left\{
\begin{array}{l}
  p + p \to \pi^0 + p + p \\
  \pi^0 \to 2\gamma
\end{array}
\right.
\end{equation}


The $\gamma$-rays emitted in opposite directions have an energy of $\sim 70$ MeV in the $\pi^0$ rest frame. The threshold energy (kinetic energy of a proton colliding with another stationary proton) for $\pi^0$ production is 280 MeV. 
Therefore, the $\gamma$-rays from the $\pi^0$ decay have a typical signature in the spectrum called "pion-decay bump" \cite{2018A&A...615A.108Y}. The identification of the "pion-decay bump" would allow us to sign the underlying hadronic interactions. $\gamma$-rays, which are neutral messengers, travel in a straight line through the Galaxy until they reach Earth. In contrast, charged CRs are deviated by Galactic magnetic fields and lose the knowledge of their original \textcolor{red}{directions}. Therefore, the observation of these $\gamma$-rays allows us to point back to the source of the hadronic interaction and, consequently, to the CR accelerator.\\

In addition to the PD-component, high-energy $\gamma$-rays are also produced by electrons through synchrotron, bremsstrahlung, and inverse Compton (IC) processes. The various processes can be identified by using multi-wavelength (MWL) analysis. In a recent study by \cite{2023JCAP...04..027S}\textcolor{red}{,} such MWL analysis was performed for several Galactic SNRs. An important hadronic contribution was observed for many of them, in particular for RX J1713.7-3946 and HAWC J2227+610. These sources were selected to be the objects of this study.\\


High-energy $\gamma$-rays interact with the atmosphere, generating cascades of particles inducing Cherenkov light. This light can be detected by ground-based Cherenkov telescopes. 
The current generation of Imaging Atmospheric Cherenkov Telescopes (IACTs) such as H.E.S.S. \cite{Aharonian_2006}, the Major Atmospheric Gamma-Ray Imaging Cherenkov telescopes (MAGIC) \cite{Albert_2008} and the Very Energetic Radiation Imaging Telescope Array System (VERITAS) \cite{Weekes_2002} have provided a wealth of information on objects emitting high-energy $\gamma$-rays.
Currently, the next generation instrument Cherenkov Telescope Array Observatory (CTAO) is being constructed \cite{2019scta.book.....C}. CTAO will consist of more than 100 telescopes, distributed between two sites: La Palma (Canary Island), in the northern hemisphere, and near the Cerro Paranal Observatory in Chile, in the southern hemisphere, thereby covering the entire night sky. CTAO will cover $\gamma$-ray energies from 20 GeV to 300 TeV, thanks to its three types of telescopes: the Small-Sized Telescope (SST), the Medium-Sized Telescope (MST) and the Large-Sized Telescope (LST). In its Alpha configuration, that includes 14 MSTs and
37 SSTs in the southern site and 4 LSTs, and 9 MSTs in the northern site, CTAO will provide an energy resolution below 10\% for energies larger than 1 TeV, an angular resolution better than 0.1\textdegree, and a sensitivity 10 times greater than its predecessors.\\

The goal of this paper is to study the potential of CTAO to distinguish protons from heavy CRs through observations of $\gamma$-rays and to obtain information on different hadronic acceleration scenarios. Two SNRs have been considered: RX J1713.7-3946 and HAWC J2227+610. To achieve this, we have considered different compositions of GCRs and different shapes of particle distributions predicted by acceleration models. Simulations for CTAO were performed by using radiative models with the parameters obtained by the MWL study \cite{2023JCAP...04..027S}. 


\section{Selection of the SNR sources}

\subsection{MWL study of Galactic SNRs}

The MWL analysis of 9 Galactic SNRs was realized by \cite{2023JCAP...04..027S}. This analysis assumed that the observed spectral energy distribution (SED) was due to the synchrotron, bremsstrahlung, IC and PD processes. A lepto-hadronic (synchrotron, bremsstrahlung, IC and PD processes) scenario, and a pure leptonic scenario (synchrotron, bremsstrahlung, IC processes) were taken into consideration to model the observed flux. Based on a likelihood comparison between \textcolor{red}{the pure} leptonic and lepto-hadronic scenarios, it was shown that the $\gamma$-ray spectra of RX 1713.7-3946 and HAWC J2227+610 were more accurately reproduced when the lepto-hadronic scenario with a predominant hadronic contribution at high energies, was considered. Moreover, at an energy of 100 TeV, RX J1713.7-3946 has a flux of $5\times10^{-12}$\,erg\,cm$^{-2}$\,s and HAWC J2227+610 has a flux of $5\times10^{-13}$\,erg\,cm$^{-2}$\,s, both of which are above the CTAO sensitivity (for an observation duration of 50 hours).



\subsection{The selected sources}
\label{sectionMWL}
\subsubsection{RX J1713.7-3946}

RX J1713.7-3946 (also known as \textcolor{red}{HESS J1713-397}, radio designation G347.3-0.5) is the most-studied young $\gamma$-ray shell SNR \cite{2004Natur.432...75A}. It was discovered in the ROSAT all-sky survey \cite{pfeffermann1996int} and is estimated to be at a distance of 1 kpc \cite{2003PASJ...55L..61F}. The source is consistent with a core-collapse SNR (type II), for which the stellar winds from the high-mass progenitor evacuated the surrounding matter prior to the SN explosion \cite{Sano_2021}, which is also suggested by observations \cite{Slane_1999}.
Correlation studies between the interstellar gas, X-ray and $\gamma$-ray emissions provide evidence for hadronic $\gamma$-ray emission in addition to the broadband emission spectra \cite{2012ApJ...746...82F}. These results are in agreement with the hadronic component observed in the MWL analysis of \cite{2023JCAP...04..027S}. Furthermore, simulations reveal that with an exposure time of 50 hours, CTAO would be able to identify the dominant $\gamma$-ray emission component from the morphology study of the SNR \cite{Acero_2017}.


\subsubsection{HAWC J2227+610}

HAWC J2227+610 (also called SNR G106.3+02.7) was first discovered by the Northern Galactic Plane survey with the Dominion Radio Astrophysical Observatory (DRAO;  \cite{1990A&AS...82..113J}) and is located at a distance of 800 pc \cite{Kothes_2001}. 
To the North of the head region, there is a pulsar wind nebula (PWN) called "Boomerang". It is powered by the pulsar PSR J2229+6114 (associated with LHAASO J2226+6057, \cite{2021Natur.594...33C}) formed from the core-collapse SN explosion leading to the SNR G106.3+2.7 \cite{Fang_2022}. The VERITAS collaboration detected significant TeV $\gamma$-ray emission from the elongated radio extension of this SNR \cite{Acciari_2009}. The extended $\gamma$-ray emission  spatially coincides with molecular clouds traced by $^{12}$CO (J = 1 - 0) emission \cite{Heyer_1998, Kothes_2001}, favouring a hadronic origin of the $\gamma$-ray emission, also supported by \cite{2023JCAP...04..027S}. However, the origin of the emission remains unclear, with three possible sources: hadronic, leptonic, or a combined process. According to \cite{2022icrc.confE.904V}, CTAO would provide information on the morphology of the source, with a significant detection of the extension of the source, thereby providing clearer evidence on the origins of the $\gamma$-ray emission.\\ 

\noindent
The global parameters of RX J1713.7-3946 and HAWC J2227+610 are detailed in Table \ref{table1}.

\begin{table*}
\centering
\captionsetup{}
\caption{The characteristics of the studied SNRs, together with the references for the data. Shell: most of the radiation comes from a shell of shocked material. Int.: SNR is interacting with the surrounding medium.}
\begin{tabularx}{\textwidth}{XXX}
    \hline
    Sources & RX J1713.7-3946 & HAWC J2227+610\\
    \hline
    Type & Shell & Int.\\
    Distance (kpc) & 1 & 0.8\\
    Galactic coordinates (deg) & (347.34,-0.47) & (106.58,2.91)\\
    High-energy $\gamma$-ray data  & ASDC/\textcolor{red}
    {Suzaku\cite{Tanaka_2008}/\textit{Fermi}
    \cite{Abdo_2011}/H.E.S.S
    \cite{2007A&A...464..235A}} & ASDC/Chandra/\textit{Fermi-LAT}\textcolor{red}{\cite{{2019ApJ...885..162X}}, VERITAS\cite{Acciari_2009}, LHAASO\cite{2021Natur.594...33C} and HAWC\cite{Albert_2020}}\\
    \hline
\end{tabularx}
\label{table1}
\end{table*}

\section{Evolution of the particle spectrum}

\subsection{Composition of the CRs}
\label{section31}

\begin{figure*}
\begin{center}
    \includegraphics[scale=0.6]{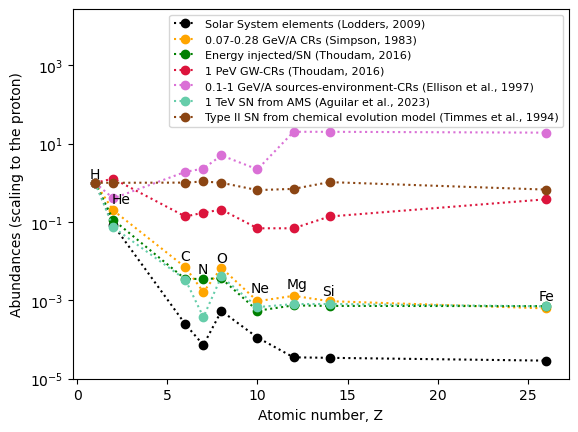}
    \caption{
    Different CR compositions. See text.}
    \label{figure1}
\end{center} 
\end{figure*}

\noindent
As discussed in Section \ref{intro}, the SNRs are believed to accelerate CRs up to the "knee" energies due to their large energy budget. In the simple Hillas formalism, the maximum energy for CRs obtained in the acceleration mechanism is proportional to the charge of the nucleus \cite{1984ARA&A..22..425H}. Therefore, heavier nuclei with charge $Z$ get accelerated to energies $Z$ times larger. This seems to be confirmed by the fact that the chemical composition at the knee shows a transition from light to heavier nuclei, from the "proton knee" at 3 PeV to the "iron knee" at about 90 PeV. In the following, we discuss several predictions and measurements of the CR composition at various energies. These compositions are shown in Figure \ref{figure1}.\\

\noindent
The composition \textcolor{red}{of} CRs measured at around 1 GeV is mainly protons (89\%), including helium nuclei (10\%) and heavier nuclei (1\%) \cite{2014ChPhC..38i0001O}. 
The CR composition measured by the Alpha Magnetic Spectrometer (AMS) detector 
 at 1 TeV \cite{PhysRevLett.130.211002} (\textit{1 TeV SN from AMS}, Figure \ref{figure1}) and measured at around 1 GeV \cite{1983ARNPS..33..323S} (\textit{0.07-0.28 GeV/A CRs}, Figure \ref{figure1}) was found to be characterized by a general overabundance of light elements such as protons and He. \cite{Thoudam_2016} also used this GCR composition at around 1 GeV to consider the amount of energy channelled into CRs by a single SN event (\textit{Energy injected/SN}, Figure \ref{figure1}).\\

\noindent
With increasing energy, the proportion of heavy nuclei becomes important (\textit{1 PeV GW-CRs} in Figure \ref{figure1}) for a CR composition at 1 PeV. This energy would be available if we consider the re-acceleration of the CRs by the Galactic wind termination shock of the SNR \textcolor{red}{\cite{Thoudam_2016}}.\\

\noindent
On the other hand, the composition is related to the type of the source. In the case of a core-collapse SN (Type II) such as RX J1713.7-3946 or HAWC J2227+610, an important contribution of heavy elements, including CNO, Ne, Mg, Si and Fe, is observed \cite{1995ApJS...98..617T} (\textit{Type II SN from chemical evolution model} in Figure \ref{figure1}). Primary CRs (He, CNO, Ne, Mg, Si, Fe,...) are synthesised in stars and accelerated by astrophysical sources. The Fe nuclei are thought to be synthesized mainly in core-collapse SNs of massive stars. Its radioactive lifetime of 3.8 Myr is sufficiently long, so that it can potentially survive the time interval between the nucleosynthesis and the detection at Earth \cite{10.1093/mnras/stab2533}.\\

\noindent
Finally, even at low energies (around 1 GeV), the CR composition can exhibit a general overabundance of heavy elements relative to protons and helium due to the interaction of CRs with the gas and dust environment \cite{Ellison_1997} (\textit{0.1-1 GeV/A sources-environment-CRs} in Figure \ref{figure1}).\\

\noindent
In Figure \ref{figure1}, one can also see the elemental composition of the Solar System \cite{2009LanB...4B..712L} (\textit{Solar System elements}), which is different from the CR composition due to the propagation effects of CRs.\\

\noindent
The two SNRs considered in this study are both core-collapse SNRs. Therefore, we have considered the composition of type II SN. In addition, we have considered GCR composition measured at 1 PeV. These two CR compositions are indicated by the brown and red points in Figure \ref{figure1}.

\subsection{Acceleration models}
\label{proton_acc}

\noindent
The differential energy distribution of accelerated hadrons, n(E), is in the following assumed to follow a simple power-law with spectral index $\alpha$ and an exponential cutoff at an energy $E_{\text{cutoff}}$, with sharpness
described by the parameter $\beta$: $n(E) \propto E^{-\alpha} e^{-(\frac{E}{E_{\text{cutoff}}})^{\beta}}$.

\noindent
Studying the shape of the energy spectra of CR particles gives us information on the underlying acceleration mechanisms.
Various physical processes, such as absorption by the source or interactions with \textcolor{red}{its} surrounding environment, can potentially influence this particle spectrum shape. According to \cite{Gabici_2014}, the CRs accelerated at the SNR, which originated from a massive star in a molecular cloud, diffusively penetrate the dense clumps that survive inside the SNR. This implies that the spectrum of the CRs inside the clumps may well be significantly harder than the one accelerated at the shock \cite{1996A&A...309..917A, 2007Ap&SS.309..365G}.\\

\noindent
According to \cite{Ang_ner_2023}, the exact shape of the cut-off in the spectrum depends, in principle, on what is limiting the acceleration. Assuming that the main mechanism responsible for CR acceleration is DSA \cite{Cristofari_2021}, the most stringent constraint is usually imposed by the size of the accelerator compared to the diffusion distance of the highest energy particles. The diffusion coefficient D can be expressed as $D(E)=D_0E^{\beta}$, where $D_0$ is a normalisation coefficient and $\beta$ is related to the sharpness of the cutoff ($\beta$ in Eq. \ref{equation2}).
In particular, an exponential cut-off is found for Bohm diffusion ($\beta = 1$), while sub-exponential cut-offs result from other diffusion models commonly used in astrophysics, such as Kolmogorov's ($\beta  = 1/3$) or Kraichnan's ($\beta  = 1/2$) models. It is expected that SNRs can reach energies close to the "knee" energy only by considering the so-called non-resonant streaming instability \cite{2004MNRAS.353..550B}, which is induced by the particles at the instantaneous maximum energy leaving the accelerator. In this scenario, the spectrum at the shock is usually assumed to be cut very sharply at $E_{\text{max}}$, which would reflect the case of super-exponential cut-offs ($\beta > 1$).\\

\noindent
Regarding RX J1713.7-3946, it is expected that the origin of the $\gamma$-ray emission would be hadronic if the SNR expands in a clumpy medium \cite{Inoue_2012, 2010ApJ...708..965Z}. 
The presence of clumps can explain the observed $\gamma$-ray spectrum of RX J1713.7-3946, which is generally considered too hard to be explained by hadronic interactions. According to simulations, the interaction between the SNR and the clumps would lead to a strong amplification of the turbulent magnetic field at the interface between the clumps and the diffuse gas. High-energy CRs would penetrate the clumps, exhibiting a faster diffusion. Consequently, the CR spectrum within the clumps is expected to be considerably harder than that of CRs in the SNR shell \cite{2012IAUS..279..335I}. \\

\noindent
These arguments led us to consider proton acceleration models with different diffusion scenarios with $\beta = 0.5$ to $\beta = 1.5$.

\section{Modelling and simulation of the $\gamma$-ray spectra for CTAO}

\subsection{Modelling with \textsc{Gammapy}}
The modelling of the $\gamma$-ray spectra were exclusively conducted using the \textsc{Gammapy} package \cite{2019A&A...625A..10N}. \textsc{Gammapy} is an open-source analysis package developed to facilitate high-level analysis of IACT data. It includes tools to perform Markov Chain Monte Carlo (MCMC) fitting of radiative models to X-ray and $\gamma$-ray spectra using \textit{emcee}, an affine-invariant ensemble sampler for MCMC \cite{Foreman_Mackey_2013}. We used the \textsc{Naima} package \cite{naima} to build our models based on the non-thermal radiation from relativistic particle populations. For more details, see Section \ref{sectionGammapy}.

\subsection{$\gamma$-ray spectrum modelling}

\subsubsection{Particle spectrum by SNR}
\label{section411}
For a strong shock, such as in the case of SNRs, the particle prediction from shock acceleration theory gives a universal power law spectrum of accelerated particles, expressed as $Q_{\text{CR}}(E) \propto E^{-\alpha}$, where the spectral index $\alpha$ is typically set to 2 \cite{1983RPPh...46..973D}.\\

\noindent
The particle energy distribution function or particle function or particle distribution $n(E)$ is expected to follow an exponential cut-off power-law (EPL), expressed as: 
\begin{equation}
    n(E) =  A_{\text{m}}  \left(\frac{E}{E_\text{0}}\right)^{-\alpha}  e^{-(\frac{E}{E_{\text{cutoff}}})^{\beta}}
\end{equation} with:
\begin{itemize}
    \item $A_{\text{m}}$: amplitude of the proton distribution ($A_{\text{m}}^{\text{p}}$) or electron distribution ($A_{\text{m}}^{\text{e}}$)
    \item $E_{\text{0}}$: reference energy typically set to 1\,TeV
    \item $\alpha$: proton spectral index ($\alpha_\text{p}$) or electron spectral index ($\alpha_\text{e}$)
    \item $E_{\text{cutoff}}$: proton cut-off energy ($E^\text{p}_{\text{cutoff}}$) or electron cut-off energy ($E^\text{e}_{\text{cutoff}}$)
    \item $\beta$: cut-off exponent/sharpness
    \label{equation2}
\end{itemize}

\noindent
The proton particle function is given by: $n_\text{p}(E) =  A_{\text{m}}^\text{p} \left(\frac{E}{E_0}\right)^{-\alpha_\text{p}}  e^{-(\frac{E}{E^\text{p}_{\text{cutoff}}})^{\beta}} $.\\
If we consider heavier nuclei, $n_\text{p}(E)$ can be modified following the approach proposed by \cite{Thoudam_2016}. This modification incorporates $f$, the fraction of the abundance of the considered CRs, 
and an increased proton energy $Z \times E^\text{p}_{\text{cutoff}}$ where $Z$ is the charge of the CR. The modified function is expressed as: 
\begin{equation}
   n^\text{Z}_\text{p}(E) =  f A_{\text{m}}^\text{p}  \left(\frac{E}{E_0}\right)^{-\alpha_\text{p}}  e^{-(\frac{E}{Z \times E^\text{p}_{\text{cutoff}}})^{\beta}} 
\end{equation}

\noindent
For the fraction of the abundance $f$, we have taken the composition at 1 PeV and the composition of the type II SN, as discussed before (see Section \ref{section31}).\\

\begin{figure*}
\begin{center}
    \includegraphics[scale=0.45]{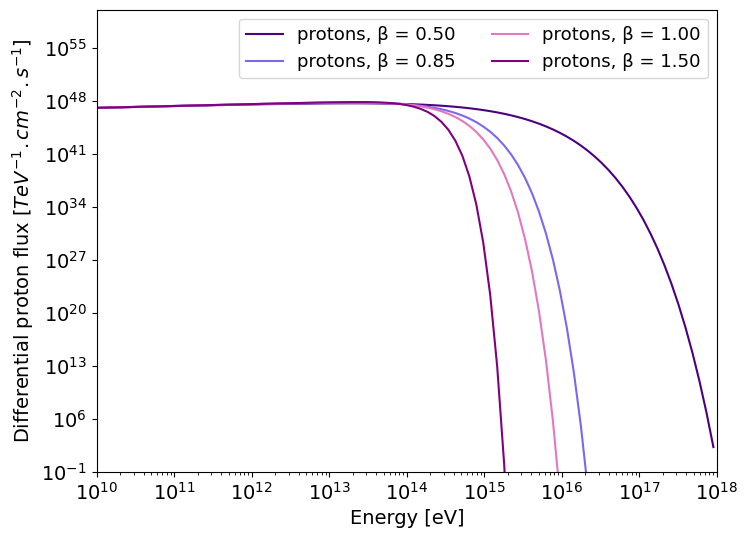}
    \caption{Proton distributions for $\beta=0.50,1.00,0.85$, and $1.50$.}
    \label{figure2}
\end{center} 
\end{figure*}

\noindent
Considering different acceleration models as mentioned in Section \ref{proton_acc}, different sharpnesses were studied. A range of values \textcolor{red}{from} $\beta = 0.50$ to $\beta = 1.50$ was considered and this study was performed only for protons for RX J1713.7-3946. Figure \ref{figure2} illustrates the proton spectrum for $\beta = 0.50$, $\beta = 0.85$, $\beta = 1.00$ and $\beta = 1.50$. The spectral parameters of the proton distribution are fixed by the MWL study of RX J1713.7-3946 \cite{2023JCAP...04..027S}.

\subsubsection{$\gamma$-ray spectrum from the $\pi^0$ decay process}

In the following, we have considered only hadronic models. As discussed before, in the case of the two studied SNRs, the hadronic component was the most important one at the highest energies. In section \ref{section54}, we compare the results obtained by using both, hadronic and leptonic (IC) models.\\

\noindent
The inelastic collisions producing $\pi^0$ are based on Glauber's multiple scattering theory \cite{2014PhRvD..90l3014K} that takes into account the inelastic interaction cross-section for protons \cite{2006PhRvD..74c4018K} and for heavier nuclei \cite{1983PhRvC..28.2178L}. The cross-section, $\sigma$, 
depends on the mass numbers of the projectile and the target, $A_{\text{pro}}$ and $A_{\text{tar}}$, the projectile and target mass number, respectively, as well as on the kinetic energy of the projectile,   $T_{\text{pro}}$. The cross-section, $\sigma$, relies on experimental data ($\pi_0$ spectra), simulations from \textit{Geant4} (interaction between particles with and surrounding matter), and hadronic models. 

\noindent
The photon flux $F^\text{Z}_\text{p}(E)$, is given by this equation:
\begin{equation}
\begin{aligned}
    F^\text{Z}_\text{p}(E) &=  f  \sigma  N_\text{H}  A_{\text{m}}^\text{p}  (\frac{E}{E_0})^{-\alpha_{\text{p}}} e^{-(\frac{E}{Z \times E^\text{p}_{\text{cutoff}} / A})^{\beta}} \\
    &=  f \sigma  N_\text{H}  A_{\text{m}}^\text{p}  (\frac{E}{E_0})^{-\alpha_{\text{p}}}  e^{-(\frac{E}{E^\text{Z}_{\text{c}}})^{\beta}}
\end{aligned}
\label{equation3}
\end{equation}

\noindent
with $\sigma$, the inelastic cross-section between GCRs and stationary protons, $N_\text{H}$, the number density of the target protons and \textit{A}, the mass number of the nuclei of the considered GCRs.\\ 

\noindent
Note that the cutoff energy in the photon spectrum for heavy CRs ($E^\text{Z}_{\text{c}}=Z \times E^\text{p}_{\text{cutoff}} / A$) is lower than that for protons ($E^\text{p}_{\text{cutoff}}$). This is due to the fact that for the $\pi^0$ creation, the relevant energy is the energy per nucleon. Thus, we expect a softening of the $\gamma$-ray spectrum in case of heavy CRs.\\



\noindent
We used the radiative PD-model from \textsc{Naima} by fixing the parameters of the proton distribution $A^{\text{p}}_{\text{m}}$, $E_{\text{0}}$, $\alpha^{\text{p}}$, $E^\text{p}_{\text{cutoff}}$, $\beta$ as well as the number density of the target protons $N_{\text{H}}$ (the density at rest is similar for protons and heavier nuclei) according to the MWL study \cite{2023JCAP...04..027S}. The fixed MWL parameters are reported in the Table \ref{table2}.

\begin{table*}
\centering
\begin{subtable}{0.65\textwidth}
\centering
	\caption{Proton distribution.}
\begin{tabular}{lcc} 
        \hline 
        & RX J1713.7-3946 & HAWC J2227+610\\
		\hline
		$A_{\text{m}}^{\text{p}}$ [TeV$^{-1}$] & \textcolor{red}{$(1.15\pm1.00)\times10^{47}$} & \textcolor{red}{$(1.05\pm1.00)\times10^{47}$}  \\
		\hline
		  $\alpha^{\text{p}}$ & $1.75\pm0.02$ & $1.76\pm0.03$\\
        \hline
		$E^\text{p}_{\text{cutoff}}$ [TeV] & $77.75\pm1.10$  & $444.16\pm1.17$\\
        \hline
		N$_{\text{H}}$ [cm$^{-3}$] & $21.03\pm1.05$ & $1.70\pm0.10$ \\
        \hline
		K$_{\text{ep}}$ [cm$^{-3}$] & \textcolor{red}{$0.0139\pm0.0006$} &  \textcolor{red}{$0.00476\pm0.00004$}  \\
        \hline
    \end{tabular}
\label{table2a}
\end{subtable}%
\hspace{0.05\textwidth} 
\begin{subtable}{0.65\textwidth}
\centering
    \caption{Electron distribution.}
\begin{tabular}{lcc} 
        \hline 
        & RX J1713.7-3946 & HAWC J2227+610\\
		\hline
		$A_{\text{m}}^{\text{e}}$ [TeV$^{-1}$] & \textcolor{red}{$(3.16\pm1.04)\times10^{46}$} & \textcolor{red}{$(2.69\pm1.08)\times10^{45}$}  \\
		\hline
		$\alpha^{\text{e}}$ &  $2.47\pm0.04$ & $1.69\pm0.03$\\
        \hline
		$E^\text{e}_{\text{cutoff}}$ [TeV] & $23.15\pm1.07$ & $0.69\pm1.00$\\
        \hline
		N$_{\text{H}}$ [cm$^{-3}$] & $21.03\pm1.05$ & $1.70\pm0.10$ \\
        \hline
		K$_{\text{ep}}$ [cm$^{-3}$] & \textcolor{red}{$0.0139\pm0.0006$} &  \textcolor{red}{$0.00476\pm0.00004$}  \\
        \hline
	\end{tabular}
\label{table2b}
\end{subtable}
\caption{Parameters obtained from the lepto-hadronic fit of the $\gamma$-ray spectrum of RX J1713.7-3946 and HAWC J2227+610 for proton and electron distributions (\textcolor{red}{from \cite{2023JCAP...04..027S}}). NB: K$_{\text{ep}}$ represents the ratio of electrons with respect to protons at 1 TeV.}
\label{table2}
\end{table*}

\noindent
We computed, with the \textsc{Naima} spectral model class, the non-thermal emission from populations of relativistic GCRs due to interactions with the surrounding matter by fixing the distance of the source under consideration (see Table \ref{table1}). \textcolor{red}{In case of composed models, such as p+CNO or p+Fe, we have combined the different individual models by using the \textsc{Gammapy} compound spectral model class.}\\

\begin{figure*}
\begin{center}
    \includegraphics[scale=0.4]{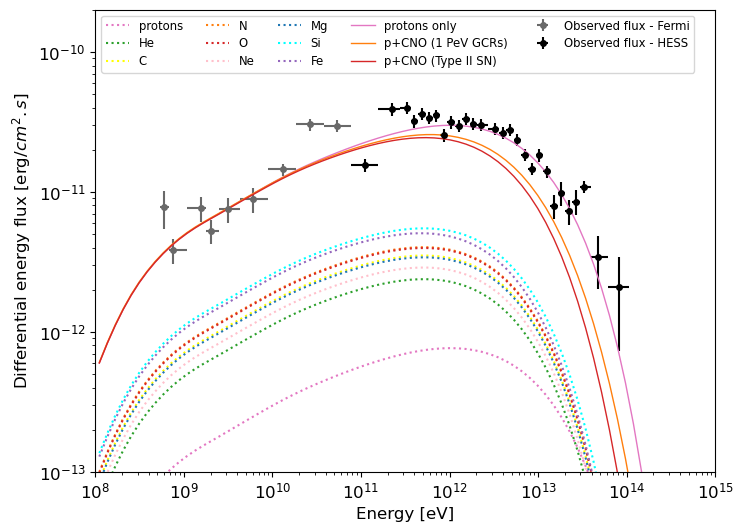}
    \includegraphics[scale=0.4]{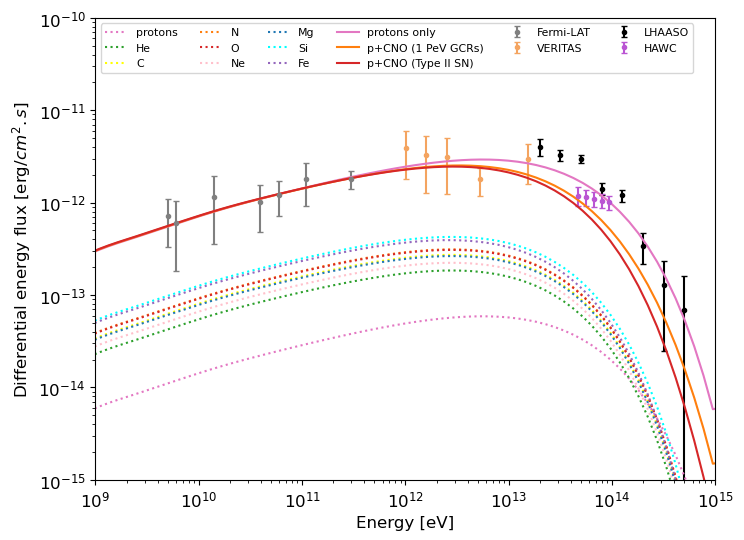}
    \caption{$\gamma$-ray spectra considering PD-models for different nuclei considering a type II SN composition (individual nuclei in dotted lines) and protons-only and composed PD-models (solid lines).
    Left: RX J1713.7-3946. Points correspond to measured data from \textit{Fermi} \cite{Abdo_2011} and H.E.S.S \cite{2007A&A...464..235A}. Right: HAWC J2227+610. Points correspond to measured data from \textit{Fermi-LAT} \textcolor{red}{\cite{{2019ApJ...885..162X}}}, VERITAS \cite{Acciari_2009}, LHAASO \cite{2021Natur.594...33C} and HAWC \cite{Albert_2020}. See text.}
    \label{figure3}
\end{center}
\end{figure*}

\noindent
The $\gamma$-ray spectra resulting from the radiative PD-models by using the particle distributions of protons and heavier nuclei (He, C, N, O, Ne, Mg, Si, Fe, p+CNO) with 1 PeV GCRs and a type II SN composition, are shown in Figure \ref{figure3}. One can observe that the different particle distributions yield different photon distributions.
Therefore, one can expect that by measuring the photon distribution with high accuracy, the underlying particle distribution can be deduced, thereby throwing light on the nature of CRs.\\
Note also that due to the global normalization, in case of 'p+CNO', the proton flux is lower than in case of 'protons only' (depending on the chosen composition, either '1 PeV GCRs' or 'type II SN').

\subsection{CTAO $\gamma$-ray spectrum simulation}

\subsubsection{Simulating with \textsc{Gammapy}}
\label{sectionGammapy}

The simulations were performed by using the \textcolor{red}{most recent version of} \textsc{Gammapy} package \cite{2023A&A...678A.157D} and the Cherenkov Telescope Array Observatory Instrument Response Functions (CTAO IRFs) available online (see Section \ref{sectiondata} - Data Availability). The CTAO IRFs are determined by Monte Carlo simulations and based on the mapping between the incoming photon flux and the detected events of the $\gamma$-ray shower.
\textsc{Gammapy} provides functionalities to reduce simulated events by binning them in energy and sky coordinates defined by a single region on the sky. Several techniques for the background estimation are implemented in the package. The counts data and the counts after the background subtraction are bundled into datasets. Additionally, \textcolor{red}{the \textsc{Gammapy} Maker class} provides the simulated data points using 
radiative models through the \textsc{Naima} package \cite{naima} (\textcolor{red}{\textsc{Naima} spectral model class}). \textcolor{red}{The spectral parameters are input parameters of the radiative models.} The data can be regrouped to compute flux points in a specific energy band \textcolor{red}{with the \textsc{Gammapy} Estimators class}. The flux of $\gamma$-ray sources can be estimated using Poisson maximum likelihood fitting.\\

\noindent
We have simulated the CTAO spectrum with a 1D On-Off analysis. This consists of measuring the spectrum of the source in a given region defined on the sky with a circular aperture centered on the region of interest ("ON-region") with a specific energy binning. To estimate the expected background in the "on region", fake off-counts are simulated with a Poisson fluctuation considering the acceptance, which is the relative background efficiency in the "OFF-region". The data were modelled using IC and/or $\pi^0$ decay models. Based on the best-fit model, the final flux points and corresponding log-likelihood profiles were computed.\\

\noindent
These simulations were performed on extended sources, as it is the case for RX J1713.7-3946 and HAWC J2227+610. 
The 1D analysis approach was chosen because RX J1713.7-3946 and HAWC J2227+610 are isolated sources. Using \href{http://gamma-sky.net/}{gamma-sky.net}, we can see that the other $\gamma$-ray sources in the region do not overlap with these sources. If there are overlapping sources or complex morphology, a 3D analysis would be necessary.

\subsubsection{Input parameters for simulations}
\label{section422}

For the simulations with \textsc{Gammapy}, we have used IRFs \textit{prod5 version v0.1} (see Section \ref{sectiondata} - Data Availability) to get the predicted number of counts. IRFs are provided for both sites (Southern or Northern), for 3 different zenith angles of the source (20\textdegree, 40\textdegree or 60\textdegree) and for 3 different observation times (0.5\,h, 5\,h or 50\,h).\\

\noindent
To run simulations, we defined the following observational parameters:
\begin{itemize}
    \item Zenith angle: the visibility plot on the \textit{TeVCat} database \cite{articletevcat} allowed us to choose the optimal zenith angle for observing the sources.
    \item Livetime: we set the observation time to 50\,h (or 200\,h). To be more realistic, we divided this lifetime into 67 (or 267) observation runs of 45\,min each, which we stacked.
    \item Offset: we set the offset to 1\textdegree. This results in a radii greater than the radii of the source, ensuring coverage of the entire source area.
    \item ON-region: this corresponds to the radii of the sources, 0.65\textdegree and 0.23\textdegree for RX J1713.7-3946 and HAWC J2227+610, respectively.
    \item Energy range: the flux was simulated in the energy range from \textcolor{red}{$3\times10^{10}$\,eV to $1.99\times10^{14}$\,eV}.
    \item Exclusion region: susceptible $\gamma$-ray sources can be localised close to the studied source. They have been determined through \href{http://gamma-sky.net/}{gamma-sky.net}. These sources do not overlap the studied sources, as explained previously.
\end{itemize}

\noindent
The following particle distributions and radiative models were used to simulate the $\gamma$-ray spectrum for RX J1713.7-3946 and HAWC J2227+610:
\begin{itemize}
    \item A composed model of protons and heavier nuclei (protons + CNO, protons + Fe) together with the radiative PD-model. 
    \item A model of protons with different sharpnesses ($\beta$=0.50, 0.85, 0.90, 0.95, 1.00, 1.05, 1.10, 1.15, 1.50) together with the radiative PD-model.
    \item A composed model of protons and heavier nuclei (protons + CNO) with the radiative PD-model and electrons with the IC radiative model (PD(p+CNO) + IC($e^-$)).
\end{itemize} 
The simulations were performed by using IC and $\pi^0$-decay radiative models with the parameters obtained from the MWL study (see Table \ref{table2}).\\

\noindent
In the case of HAWC J2227+610, the optimal zenith angle is 60\textdegree. The air attenuation is significant for high zenith angles; therefore, we decided to simulate the data at 40\textdegree. The area for collecting photons at 40\textdegree is smaller than that at 60\textdegree, resulting in lower statistics but more data points at the highest energies. In the case of RX J1713.7-3946, the optimal zenith angle is 20\textdegree, which was used for simulations. The different input parameters are indicated in Table \ref{table3}.\\

\begin{table*}
\centering
	\caption{Input parameters for CTAO simulations for RX J1713.7-3946 and HAWC J2227+610.}
\begin{tabular}{lcc} 
        \hline 
        & RX J1713.7-3946 & HAWC J2227+610\\
		\hline
		Zenith angle [deg] & 20 & 40 \\
        \hline
	    Lifetime [h] & 50 & 50 \\
		\hline
	  	  Offset [deg] & 1 & 1 \\
		\hline
   	  Site location & South: 14 MSTs / 37 SSTs & South: 14 MSTs / 37 SSTs \\
		\hline
   	  ON region radii [deg] & 0.65 & 0.23 \\
		\hline
   	  Energy range [TeV] & [\textcolor{red}{$30\times10^{-3}$},199], bin=31 & [\textcolor{red}{$30\times10^{-3}$},199], bin=31 \\
		\hline
   	  Exclusion region & CTB 37A (SNR G348.5+0.1) & SNR G106.6+2.9 \\
		\hline
		Tested models & PD: protons/p+CNO/p+Fe/p($\beta\ne1$) & PD: protons/p+CNO/p+Fe \\
        \hline
        & PD+IC: p+CNO+$e^-$ & PD+IC: p+CNO+$e^-$\\
        \hline
    	Extended source & Yes & Yes \\
        \hline
        \label{table3}
	\end{tabular}
\end{table*}

\begin{figure*}
\begin{center}
    \includegraphics[scale=0.29]{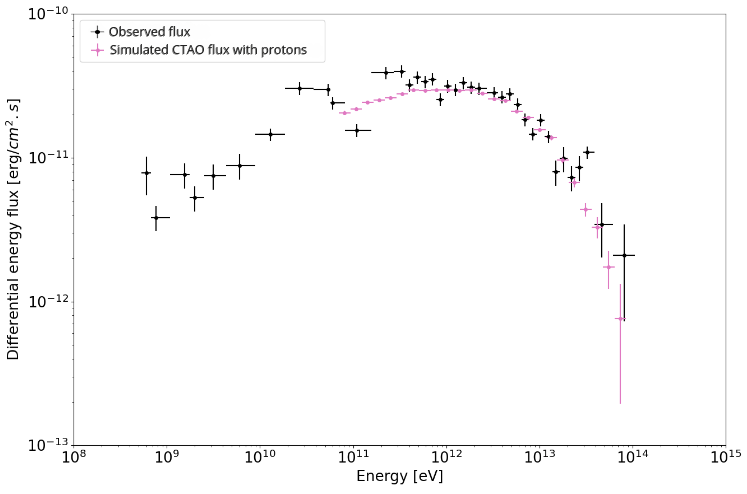}
    \includegraphics[scale=0.29]{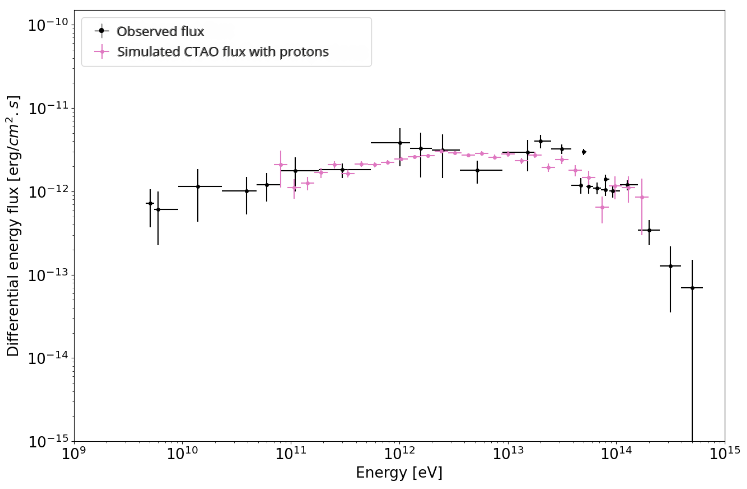}
    \caption{Left: CTAO simulation for the $\gamma$-ray spectrum of RX J1713.7-3946 using proton distribution (pink points). Black points correspond to measured data from \textit{Fermi} \cite{Abdo_2011} and H.E.S.S \cite{2007A&A...464..235A}. Right: CTAO simulation for the $\gamma$-ray spectrum of HAWC J2227+610 using proton distribution (pink points). Black points correspond to measured data from \textit{Fermi-LAT} \textcolor{red}{\cite{{2019ApJ...885..162X}}}, VERITAS \cite{Acciari_2009}, LHAASO \cite{2021Natur.594...33C} and HAWC \cite{Albert_2020}. See text.}
    \label{figure4}
\end{center}
\end{figure*}

\noindent
In Figure \ref{figure4}, we compare the CTAO simulations to currently existing observations. For the simulations, we supposed a proton distribution and a radiative PD-model shown in Figure \ref{figure3}. The data points simulated at the limits of the energy range were not considered because of the instrumental uncertainty.
As can be seen, we observe an accurate reconstruction of the CTAO flux and a good agreement with the currently existing data.
However, it is important to note that each instrument has a different level of systematic uncertainty. In the following, these existing data points have not been included in the fit procedure.

\subsubsection{Data fitting and likelihood test}

The quality of the simulation was assessed by using the \textit{Minuit} optimizer \cite{James:1975dr}, which is accessible through \textsc{Gammapy's Fit class}. The simulated fluxes were fitted using different models to determine how closely they matched the input model. A lower Loglikelihood value indicates a smaller difference between the simulated flux and the models.\\

\noindent
The log-likelihood test (TS) was realized with the maximum log-likelihood ratio $\lambda$ given by $\lambda=\mathcal{L}_1/\mathcal{L}_0$ with 0 and 1 corresponding to the two models that are compared. The $\Delta$TS value is given by:\\
$\Delta TS=TS_1-TS_0=-2ln(\lambda)$ where $TS_0$ and $TS_1$ are given by the statistics of the fit ($\text{total}_{\text{stat}}$) using a \textit{Minuit} minimizer (returning $-2\text{LogLike}$). This statistic behaves like a chi-squared function, i.e. the minimum of the function in repeated identical random trials is chi-squared distributed up to an arbitrary additive constant. Hence, $\Delta TS$ follows the $\chi^2$ distribution with $n_{\text{dof}}$ degrees of freedom, i.e. the difference in the number of free parameters between the models (0 and 1).\\

\noindent
We assumed that a significance larger than 5$\sigma$ indicates that the two models can be distinguished. This significance corresponds to $\Delta$TS equal to 25 for 1 degree of freedom and 29 for 2 degrees of freedom.

\section{Results}

\subsection{CTAO sensitivity to heavy nuclei}
\label{section5}
\begin{figure*}
\centering
    \begin{subfigure}[t]{0.45\textwidth}
        \centering
        \includegraphics[scale=0.376]{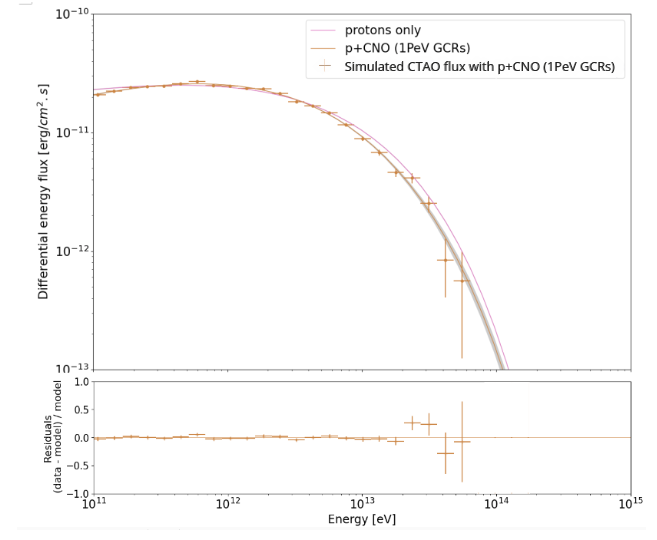}
    \end{subfigure}
    \hspace{0.08\textwidth}
    \begin{subfigure}[t]{0.45\textwidth}
        \centering
        \includegraphics[scale=0.4]{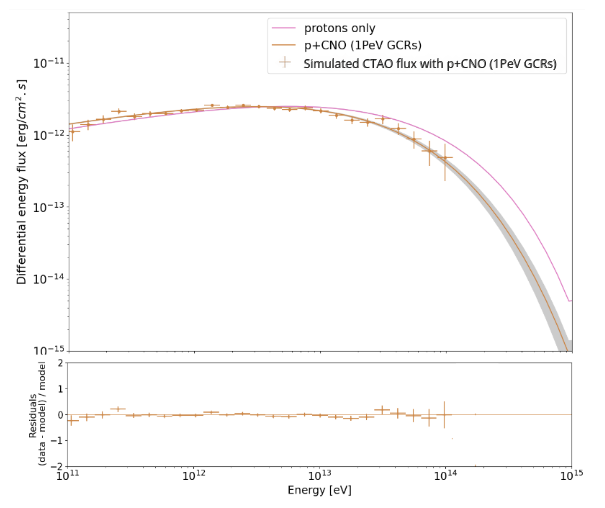}
    \end{subfigure}
    \vspace{0.5cm}
    \begin{subfigure}[t]{0.45\textwidth}
        \centering
        \includegraphics[scale=0.41]{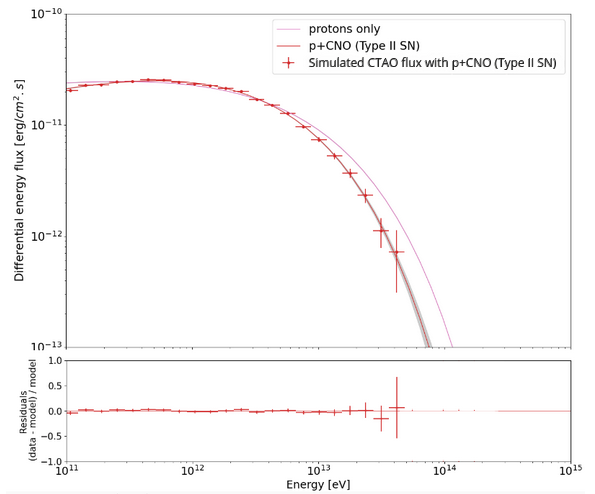}
    \end{subfigure}
    \hspace{0.08\textwidth}
    \begin{subfigure}[t]{0.45\textwidth}
        \centering
       \includegraphics[scale=0.41]{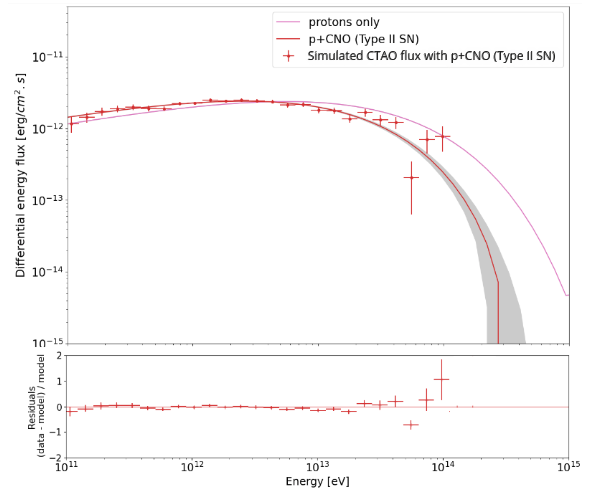}
    \end{subfigure}
    \vspace{0.5cm}
    \begin{subfigure}[t]{0.45\textwidth}
        \centering
        \includegraphics[scale=0.39]{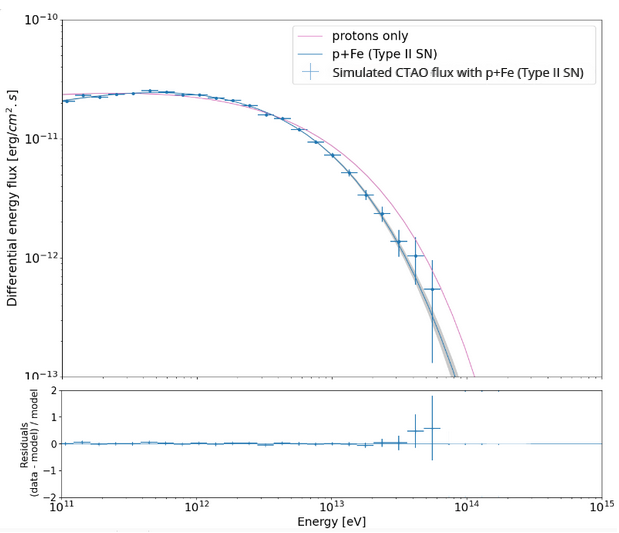}
    \end{subfigure}
    \hspace{0.08\textwidth}
    \begin{subfigure}[t]{0.45\textwidth}
        \centering
        \includegraphics[scale=0.4]{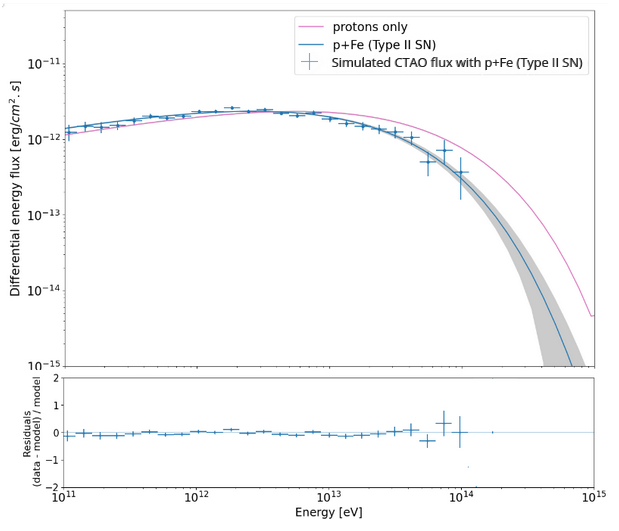}
    \end{subfigure}
    \caption{Left: CTAO simulations for the $\gamma$-ray spectrum of RX J1713.7-3946 (points) with models indicated in the figure. Right: CTAO simulations for the $\gamma$-ray spectrum of HAWC J2227+610 (points) with models indicated in the figure. The results of the fit with different models are shown in solid lines. The shaded area shows the uncertainties of the fit.}  
    \label{figure5}
\end{figure*}

This study aims to determine whether CTAO can detect heavy GCRs. To achieve this, we simulated the CTAO flux for RX J1713.7-3946 and HAWC J2227+610 using a radiative PD-model with protons-only distribution and a composed distribution of protons + CNO and protons + Fe (see Section \ref{section422}). The simulations were performed for 1 PeV GCRs and type II SN compositions. The fits were performed by using the radiative PD-model with the following free parameters for each source (see Section \ref{section411}): 
\begin{itemize}
    \item RX J1713.7-3946: amplitude $A^{\text{p}}_{\text{m}}$, spectral index $\alpha^{\text{p}}$.
    \item HAWC J2227+610: amplitude $A^{\text{p}}_{\text{m}}$.
\end{itemize}
In the case of HAWC J2227+610, the flux is approximately 150 times lower than that of RX J1713.7-3946, and the use of only 1 free parameter allowed the fit to converge.
The other model parameters were taken from the MWL study, see Table \ref{table2}.
The results are shown in Figure \ref{figure5}.\\


\noindent
Table \ref{table4a} presents the statistical results of the likelihood tests $\Delta$TS, obtained when comparing the two models. We assume that a $\Delta TS$ value greater than 25 (only 1 free parameter) or 29 (2 free parameters) indicates that CTAO can distinguish the two models. One can see that high values (>>29) were obtained for RX J1713.7-3946 in the case of using 2 free parameters. As an example, Table \ref{table4b} shows the parameters obtained from the fit of RX J1713.7-3946. These results indicate that CTAO would be capable of distinguishing between protons-only and p+CNO or p+Fe considering the two different GCRs compositions (1 PeV GCRs and type II SN).\\
Similar conclusions are obtained for HAWC J2227+610 with slightly lower statistical significance and by using only one free parameter.\\

\newpage

\begin{table*}
\centering
\begin{subtable}{0.65\textwidth}
\centering
\caption{Log-likelihood tests comparing $\gamma$-ray spectra given by heavy CRs distribution (p + CNO, p + Fe) with proton distribution. Two compositions for heavy CRs are used: 1 PeV GCRs and type II SN.}
\begin{adjustbox}{width=0.9\textwidth}
\begin{tabular}{lcc} 
        \hline 
        & RX J1713.7-3946 & HAWC J2227+610\\
		\hline
		TS[p+CNO (1 PeV GCRs)] - TS[p] & 102 & 53 \\
        \hline
    	TS[p+CNO (Type II SN)] - TS[p] & 202 &  113 \\
        \hline
    	  TS[p+Fe (Type II SN)] - TS[p] & 188  & 88 \\
        \hline
	\end{tabular}
 \end{adjustbox}
\label{table4a}
\end{subtable}%
\hspace{0.05\textwidth} 
\begin{subtable}{0.65\textwidth}
\centering
\caption{Example of the parameters obtained from the fit of RX J1713.7-3946. $A^{\text{p}}_{\text{m}}$ and $\alpha^{\text{p}}$ are set free during the fit.}
\begin{adjustbox}{width=0.9\textwidth}
\begin{tabular}{lccc} 
        \hline 
         & p+CNO (1 PeV GCRs) & p+CNO (Type II SN) & p+Fe (Type II SN) \\
		\hline
		$A^{\text{p}}_{\text{m}}$ [TeV$^{-1}$] & $(7.62\pm1.13)\times10^{46}$ & ($7.91\pm1.20)\times10^{46}$ & $(2.17\pm0.14)\times10^{47}$\\
        \hline
    	$E_{\text{0}}$ [TeV] & 1.00 & 1.00 & 1.00\\
        \hline
       	$\alpha^{\text{p}}$ &  $(1.75\pm0.06)$ & $(1.77\pm0.06)$ & $(1.76\pm0.02)$\\
        \hline
       	$E^\text{Z}_{\text{c}}$ [TeV] &  38.88 & 38.88 & 36.10\\
        \hline
       	$\beta$ &  1.00 & 1.00 & 1.00\\
        \hline
	\end{tabular}
 \end{adjustbox}
\label{table4b}
\end{subtable}
\caption{Fit results.}
\label{table4}
\end{table*}

\subsection{CTAO sensitivity to different proton acceleration models}
\label{section6}

\begin{figure*}
\begin{center}
    \includegraphics[scale=0.29]{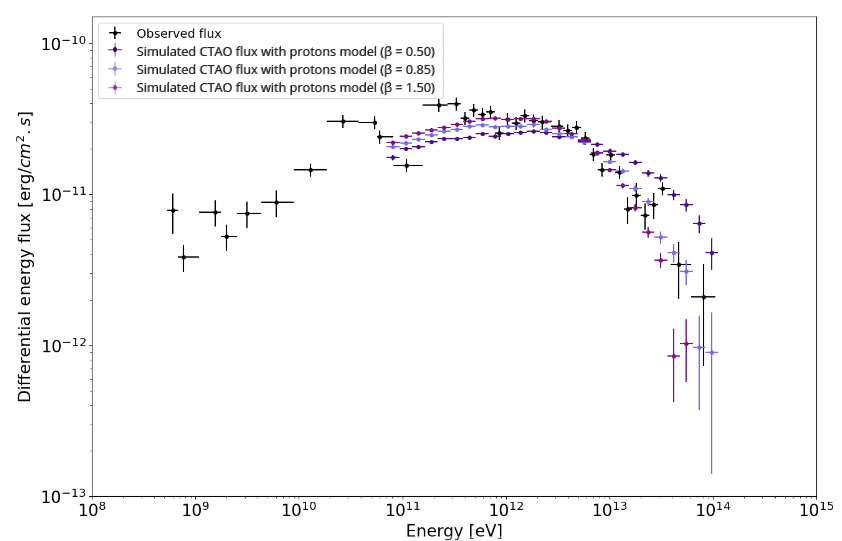}
    \includegraphics[scale=0.29]{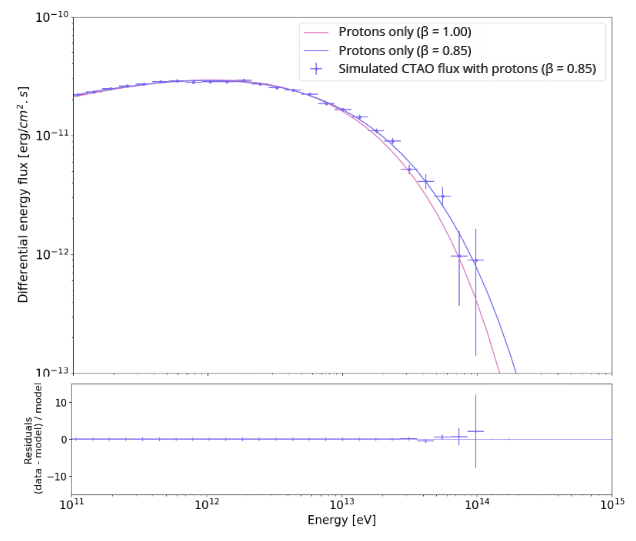}
    \caption{Left: CTAO simulations for the $\gamma$-ray spectrum of RX J1713.7-3946 by using different $\beta$ values for proton distribution (see figure insert). Black points correspond to measured data from \textit{Fermi} \cite{Abdo_2011} and H.E.S.S \cite{2007A&A...464..235A}. Right: CTAO simulation for the $\gamma$-ray spectrum by using proton distribution with $\beta=0.85$ (purple points). Fit results for proton models considering $\beta=1.00$ (pink solid line) and $\beta=0.85$ (purple solid line).}
    \label{figure6}
\end{center}
\end{figure*}

\begin{figure*}
\begin{center}
    \includegraphics[scale=0.338]{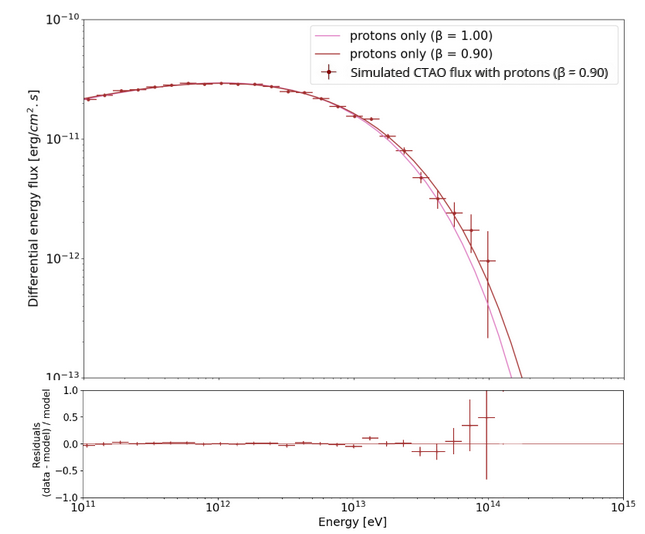 }
    \includegraphics[scale=0.338]{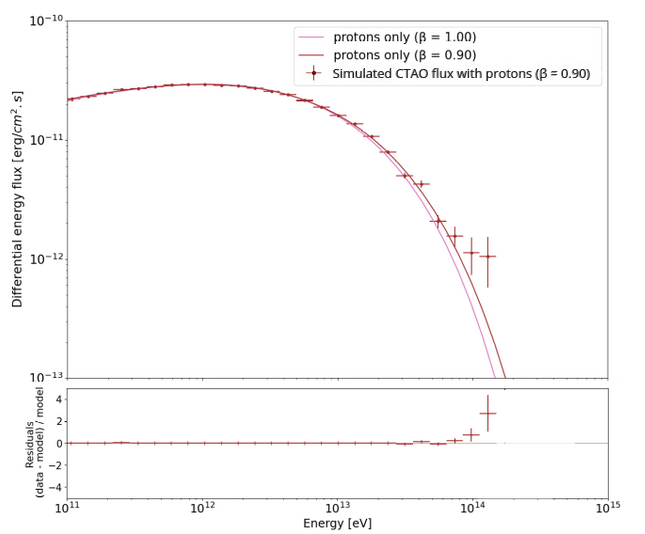}
    \caption{CTAO simulation for the $\gamma$-ray spectrum of RX J1713.7-3946 by using proton distribution with $\beta=0.90$ (brown points). Fit results for proton models considering $\beta=1.00$ (pink solid line) and $\beta=0.90$ (brown solid line). Left: results for 50\,h of observations. Right: results for 200\,h of observations.}
    \label{figure7}
\end{center}
\end{figure*}

The flux of RX J1713.7-3946 was simulated for CTAO using the PD-model for three different proton distributions with $\beta$ = 0.50, 1.50 and 0.85 (see Section \ref{equation2}). The results of the simulation are compared to measured data in Figure \ref{figure6} (left). One can observe a good agreement between measured data and simulations with $\beta$ = 0.85. It is again noted that Figure \ref{figure7} allows us only qualitative comparison; in the following, the fit procedure is performed only for simulated CTAO data.\\

\noindent
Figure \ref{figure6} (right) shows the results of the simulation with $\beta$ = 0.85 together with the fit results for $\beta$ = 0.85 and $\beta$ = 1.00.
This leads to a difference between the log-likelihood tests $\Delta TS=32$ using 2 free parameters ( $A^{\text{p}}_{\text{m}}$ and $\alpha^{\text{p}}$), showing that CTAO would be able to distinguish between these two $\beta$ values.\\

\noindent
In order to determine the CTAO sensitivity to distinguish between different $\beta$ values, we simulated CTAO data by using several $\beta$ values around 1.00, from 0.85 to 1.15, with a step of 0.05. These simulations were performed for an observation time of 50\,h and 200\,h. As an example, Figure \ref{figure7} shows the comparison of $\beta$=0.90 and $\beta$=1.00. This leads to a difference between the log-likelihood tests $\Delta TS=9$ for 50\,h and $\Delta TS=46$ for 200\,h. These results indicate that 50\,h would not be sufficient to distinguish a $\beta$ variation of 0.10. Longer observation time increases statistics, resulting in better sensitivity for different $\beta$ values. Table \ref{table5} shows a compilation of all results obtained for different $\beta$ values using 50\,h and 200\,h observation times. The statistical analysis is performed by comparing simulations with different $\beta$ values to the model of $\beta$=1.00. One can observe that for 200\,h, a $\beta$ variation as small as 0.10 can be distinguished. In the case of 50\,h, only variations larger than 0.15 can be differentiated. \\

\noindent
This study shows that CTAO is sensitive to the shape of the particle distribution and, therefore, could provide information on different acceleration models.

\begin{table*}
\centering
\caption{Log-likelihood tests comparing CTAO simulations for the $\gamma$-ray spectrum of RX J1713.7-3946 by using proton distribution with different $\beta$ values to that with $\beta$=1.00.}
\begin{tabular}{lccccccccc} 
        \hline 
       & $\beta$ & $0.50$ &  $0.85$ & $0.90$ &  $0.95$ 
       &  $1.05$ &  $1.10$ & $1.15$ & $1.50$  \\
		\hline
		  50\,h & TS[p($\beta$)] - TS[p($\beta=1$)] & 649 & 33 &  7 & 2  & 7 & 4 &  7 & 98   \\
        \hline
		  200\,h & TS[p($\beta$) - TS[p($\beta=1$)] & 2500 &  98 &  48 & 11 & 4 & 23 & 40 & 471 \\
        \hline
        \label{table5}
	\end{tabular}
\end{table*}

\subsection{CTAO sensitivity to separate heavy nuclei and proton acceleration models}

The goal of this study was to determine whether CTAO could differentiate between the contribution of heavy nuclei and different $\beta$ values. This was done by simulating CTAO fluxes for an observation time of 50\,h using a proton distribution with $\beta$=0.85 and a protons+CNO distribution with $\beta$=1.00. For the CR composition, type II SN and 1 PeV GCRs were used. \\

\begin{figure*}
\begin{center}
    \includegraphics[scale=0.338]{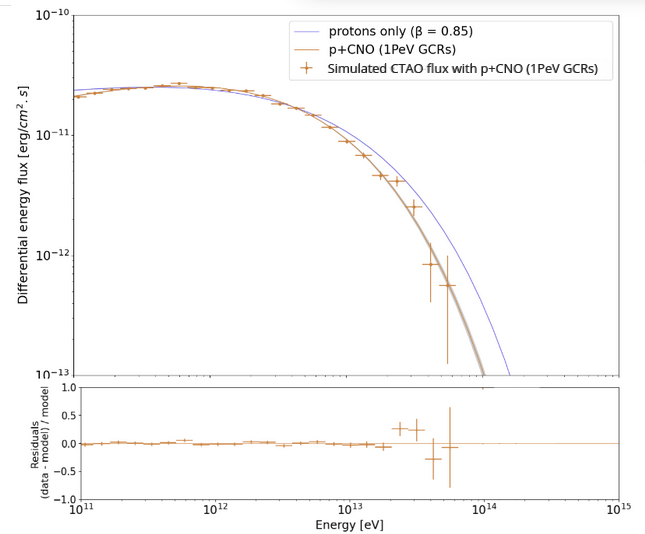}
    \includegraphics[scale=0.338]{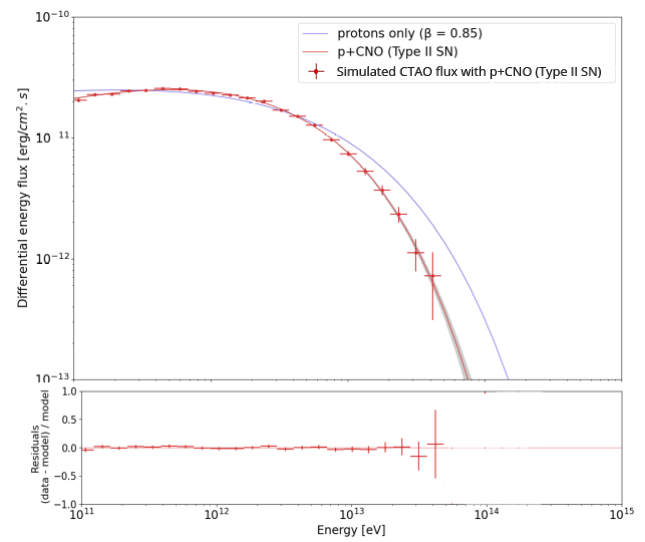}
    \caption{Left: CTAO simulation for the $\gamma$-ray spectrum of RX J1713.7-3946 by using a p+CNO distribution with a 1 PeV composition (orange points). Fit results for a proton distribution with $\beta$=0.85 (blue solid line). Right: CTAO simulation for the $\gamma$-ray spectrum of RX J1713.7-3946 by using p+CNO distribution with a type II SN composition (red points). Fit results for a proton distribution with $\beta$=0.85 (blue solid line).} 
    \label{figure8}
\end{center}
\end{figure*}

\noindent
The results are shown in Figure \ref{figure8}. Table \ref{table6} compiles all obtained results. As can be seen, $\Delta$TS values for all combinations are greater than 29, indicating that CTAO can distinguish between different CR compositions and $\beta$ values. This is due to the shape of the $\gamma$-ray spectrum, which is different for heavy nuclei compared to the proton spectrum with different $\beta$ values. Therefore, we can conclude that the high resolution and the sensitivity of CTAO would allow us to distinguish between the contribution of heavy nuclei and different $\beta$ values. However, to do such studies, it is necessary to perform MWL analysis \cite{2023JCAP...04..027S} in order to fix some of the spectral parameters.\\



\begin{table*}
\centering
\caption{Log-likelihood tests comparing $\gamma$-ray spectra given by heavy CRs distribution (p + CNO) with proton distribution ($\beta$=0.85) for RX J1713.7-3946. Two compositions for heavy CRs are used: 1 PeV GCRs and type II SN.}
\begin{tabular}{lccc} 
        \hline
        Composition for p+CNO & 1 PeV GCRs & Type II SN\\
	    \hline
		TS[p+CNO] - TS[p ($\beta=0.85$)] & 191 & 317\\
         \hline
        \label{table6}
	\end{tabular}
\end{table*}

\noindent
\textcolor{red}{We also studied the use of \textsc{Naima} Exponential Cutoff Broken Power Law (EBPL) distribution instead of EPL for CRs. The results show that there is no significant change in the conclusions. This can be understood since the shape of the spectrum induced by adding heavy nuclei is clearly different from other types of cutoff shapes (different $\beta$ values or using different
distributions, EPL or EBPL).}

\newpage

\subsection{Contribution of electrons}
\label{section54}
Up to now, we have considered only hadronic scenarios with the PD-model. As discussed before, hadronic contribution dominates at the highest energies in the case of the two studied sources. Finally, to study the possible impact of electrons on the previous conclusions, we have included electron distribution and the IC radiative model by using the parameters from the MWL study (see Table \ref{table2}). The parameters for the electron distributions are mostly fixed by the synchrotron and bremsstrahlung contributions at low energies. \\

\begin{figure*}
\begin{center}
    \includegraphics[scale=0.32]{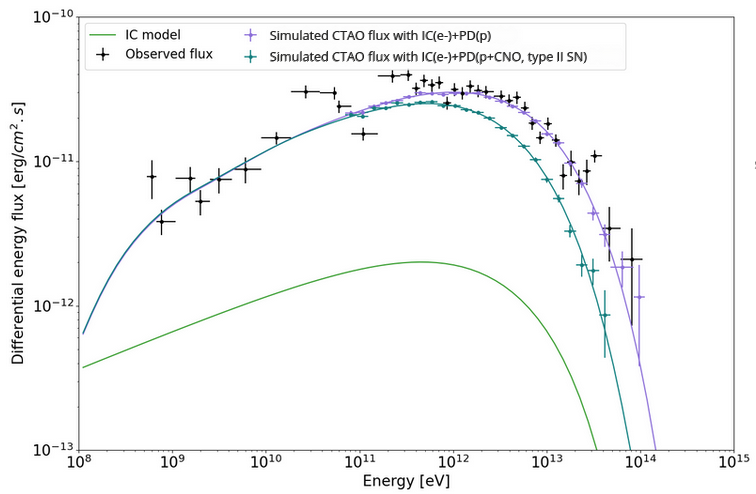}
    \includegraphics[scale=0.33]{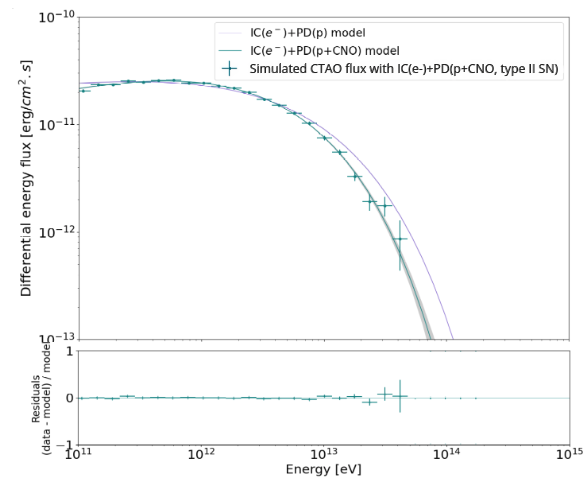}
    \caption{Left: CTAO simulations for the $\gamma$-ray spectrum of RX J1713.7-3946 by using PD+IC-model with proton distribution (purple points) and PD(p+CNO)+IC-model with p+CNO distribution (blue points, see figure insert). The solid green line corresponds to the IC contribution. Black points correspond to measured data from \textit{Fermi} \cite{Abdo_2011} and H.E.S.S \cite{2007A&A...464..235A}. Right: CTAO simulation for the $\gamma$-ray spectrum by using PD(p+CNO)+IC-model considering a type II SN composition (blue points). Fit results for the PD+IC-model (purple solid line) and for the PD(p+CNO)+IC-model (blue solid line).}
    \label{figure9}
\end{center}
\end{figure*}

\begin{figure*}
\begin{center}
    \includegraphics[scale=0.34]{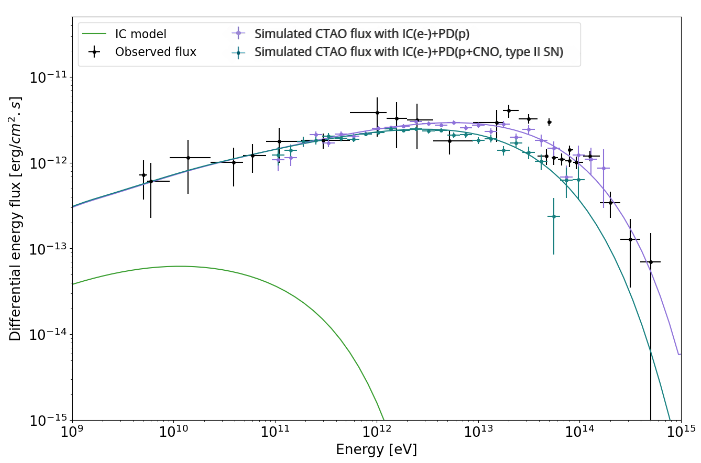}
    \includegraphics[scale=0.28]{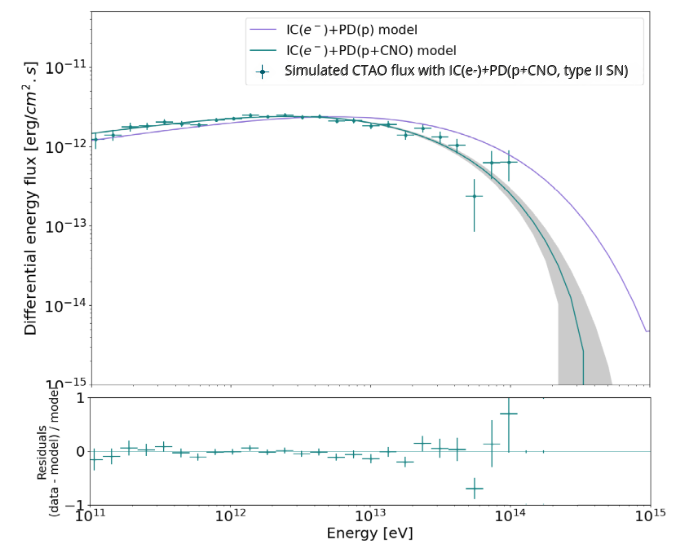}
    \caption{Left: CTAO simulations for the $\gamma$-ray spectrum of HAWC J2227+610 by using a PD+IC-model with proton distribution (purple points) and PD(p+CNO)+IC with p+CNO distribution (blue points, see figure insert). The solid green line corresponds to the IC contribution. Black points correspond to measured data from \textit{Fermi-LAT} \textcolor{red}{\cite{{2019ApJ...885..162X}}}, VERITAS \cite{Acciari_2009}, LHAASO \cite{2021Natur.594...33C} and HAWC \cite{Albert_2020}. Right: CTAO simulation for the $\gamma$-ray spectrum by using PD(p+CNO)+IC-model considering a type II SN composition (blue points). Fit results for the PD+IC-model (purple solid line) and for the PD(p+CNO)+IC-model (blue solid line).} 
    \label{figure10}
\end{center}
\end{figure*}

\noindent
We have simulated CTAO flux by using a composed model of PD(p+CNO)+IC and fitted it with a composed model of PD(p)+IC. A composition of a type II SN for p+CNO nuclei was used and $\beta$=1.00 was taken for both models. The parameters $A^{\text{p}}_{\text{m}}$ and $\alpha^{\text{p}}$ for PD-models were kept free during the fit for RX J1713.7-3946. For the fit for HAWC J2227+610, only $A^{\text{p}}_{\text{m}}$ for PD-models was kept free. The results are shown in Figure \ref{figure9} for RX J1713.7-3946 and in Figure \ref{figure10} for HAWC J2227+610. The $\Delta$TS test gave high values: 180 for RX J1713.7-3946 and 105 for HAWC J2227+610 (see Table \ref{table7}). These results show that the inclusion of electron distribution and IC model still \textcolor{red}{allow} us to distinguish between protons and p+CNO. This is consistent with the analysis realized in Section \ref{section5} by using only hadronic models.
However, a MWL analysis is necessary to fix the parameters of the IC-model.





\begin{table*}
\centering
\caption{Log-likelihood tests comparing $\gamma$-ray spectra given by heavy CRs and electron distributions (p + CNO + e$^-$) with proton and electron distributions (p + e$^-$). The type II SN composition for heavy CRs was used.}
\begin{tabular}{lccc} 
        \hline
        Source & RX J1713.7-3946 & HAWC J2227+610 \\
	    \hline
		TS[p+CNO+e$^-$] - TS[p+e$^-$] & 180 & 105\\
         \hline
        \label{table7}
	\end{tabular}
\end{table*}

\section{Conclusions}

We have studied the impact of CTAO on the spectral shape sensitivity in the cases of two SNRs: RX J1713.7-3946 and HAWC J2227+610. These SNRs were previously studied in a MWL analysis that allowed us to fix some of spectral parameters. The spectrum shape was modified either by the inclusion of heavy CRs or by changing the sharpness $\beta$ of the cut-off of the accelerated particle spectrum.\\


\noindent
In the first study, the spectrum shape was changed by the inclusion of heavy CRs, in particular, CNO and Fe, for which we have taken two compositions: type II SN and GCRs composition at 1 PeV. We observed a significant softening in the spectra of the $\gamma$-ray sources which is consistent with the conclusions of \cite{2022A&A...661A..72B}.
For both compositions, we showed that CTAO will be able to distinguish between the $\gamma$-ray spectrum measured for protons and for protons+CNO or protons+Fe. CTAO would, therefore, help us to point back to GCR accelerators and to understand the sources responsible for the proton and Fe knee that is observed in the CR spectrum.\\

\noindent
The diffusion of protons at the CR source has been described using different acceleration models. Different diffusion coefficients are related to the sharpness $\beta$ of the particle distribution cut-off at high energies. We have performed CTAO simulations for RX J1713.7-3946 by using $\beta$ values ranging from 0.50 to 1.50. The results showed that CTAO would be sensitive to a minimal change in $\beta$ values of 0.10 and 0.15 considering 200\,h and 50\,h observation times, respectively. This sensitivity would allow CTAO to provide information on the different proton acceleration models. Furthermore, we have shown that the different shapes of the spectrum induced by heavy CRs and the change of $\beta$ would allow CTAO to also distinguish between these two scenarios.\\

\noindent
This study concerned Galactic SNRs. The CTAO Galactic Plane Survey (GPS) \cite{Remy_2021} has shown that Pulsar Wind Nebulae (PWNe) and SNRs are the two dominant Galactic $\gamma$-ray sources that would be detected by CTAO. The survey yielded over two hundred PWNe and several tens of SNRs \cite{consortium2023prospects}. Therefore, CTAO has the potential to double the number of SNRs observed at TeV energies. Its high angular and energy resolution would allow us to perform detailed studies on a large number of Galactic sources and get information on the $\gamma$-ray sources of GCRs and their accelerators.

\section*{Acknowledgements}

We gratefully acknowledge financial support from the agencies and
organizations listed here: \url{http://www.ctao-observatory.org/consortium_acknowledgments}. This paper has gone through internal review by the CTAO Consortium.

\section*{Data Availability}
\label{sectiondata}
This research has made use of the CTAO instrument response functions provided by the CTAO Consortium
and Observatory, see \url{https://www.ctao-observatory.org/science/ctao-performance/} (version prod5 v0.1; \cite{cherenkov_telescope_array_observatory_2021_5499840}) for more details. Available online: \url{https://zenodo.org/record/5499840\#.YfIV5fgRVPY}.




\bibliographystyle{JHEP}
\bibliography{references} 







\label{lastpage}
\end{document}